\documentclass[12p]{article}
\usepackage[T1]{fontenc}
\usepackage{graphicx}
\usepackage{subfig}
\usepackage{amsmath,amssymb,amsfonts}
\usepackage{enumitem}
\usepackage{multirow}
\usepackage{mathrsfs}
\usepackage{url}

\date{}
\begin{document}
\title{A Lightweight Approach for User and Keyword Classification in Controversial Topics}
%
%
\author{Ahmad Zareie, Kalina Bontcheva, Carolina Scarton}

\maketitle

{\centering Department of Computer Science, The University of Sheffield, Sheffield, UK \\
\{\textit{a.zareie, k.bontcheva, c.scarton}\}\textit{@sheffield.ac.uk}}

\maketitle              
\begin{abstract}
Classifying the stance of individuals on controversial topics and uncovering their concerns is crucial for social scientists and policymakers. Data from Online Social Networks (OSNs), which serve as a proxy to a representative sample of society, offers an opportunity to classify these stances, discover society's concerns regarding controversial topics, and track the evolution of these concerns over time. Consequently, stance classification in OSNs has garnered significant attention from researchers. However, most existing methods for this task often rely on labelled data and utilise the text of users' posts or the interactions between users, necessitating large volumes of data, considerable processing time, and access to information that is not readily available (e.g. users' followers/followees). This paper proposes a lightweight approach for the stance classification of users and keywords in OSNs, aiming at understanding the collective opinion of individuals and their concerns. Our approach employs a tailored random walk model, requiring just one keyword representing each stance, using solely the keywords in social media posts. Experimental results demonstrate the superior performance of our method compared to the baselines, excelling in stance classification of users and keywords, with a running time that, while not the fastest, remains competitive.

\end{abstract}
\section{Introduction}
In society, individuals may hold varying opinions on controversial topics. Controversial topics refer to subjects that create polarisation in individuals' stances, leading to dichotomous opinions (e.g., Leave vs. Remain in the Brexit context, Pro-vaccine vs. Anti-vaccine in the COVID-19 context, or support for a specific candidate in an election); this resulting position is known as the stance.
In controversial topics, classifying users based on their stance, discovering the concerns within each class and tracking the evolution of classes and concerns over time are regarded as crucial for social scientists and policymakers~\cite{ALDAYEL2021102597}. Data collected from Online Social Networks (OSNs) 
provide valuable insights for this crucial task since it can be seen as a proxy for opinions in societies. In these networks, individuals express and discuss their viewpoints to disseminate their opinions and potentially influence others' perspectives. Therefore, the stance of users can be inferred based on their activities in OSNs; additionally, the keywords they use in their posts can express their concerns. In this paper, we propose a \textbf{random walk approach} to classify users and keywords in OSNs based on their \textbf{stance on a controversial topic}. This approach is lightweight, relying solely on keywords shared by users for the classification task. This approach only requires one keyword to represent each stance as prior knowledge, without the need for pre-determined thresholds. Hereafter, we use 'hashtag' interchangeably with 'keyword', but the method can be used with 
any token as keywords. 

Analysing data in OSNs to discover the stance of users and the generated content (posts) has increasingly become the focus of researchers from various disciplines, and a multitude of methods have been proposed for this task~\cite{ALDAYEL2021102597,alturayeif2023systematic}. 
Previous work focuses mainly on supervised models, which rely on a set of data known as labelled (training) data as prior knowledge. These supervised methods could be divided into two groups: (i) content-based~\cite{10.1145/3003433,10.1145/3110025.3110112}: requiring a set of texts as labelled data and utilising the text of posts to understand the stance, and (ii) interaction-based~\cite{10.1145/3359307,10068608,zhou2023stance}: relying on a set of users representing each stance as labelled data and using the interactions between pairs of users (e.g., likes, retweets, replies, and mentions on Twitter -- now X) to understand the stance. The labelled data required in supervised methods is typically generated through a manual process, involving human judgments, which is time-consuming (particularly when tracking the evolution of public opinion over time~\cite{ALDAYEL2021102597}) and highly affected by annotators' biases. Data labelling may also lead to language- and domain-dependant models, requiring further data annotation for handling new domains and languages. 
Unsupervised methods~\cite{Darwish_Stefanov_Aupetit_Nakov_2020,kobbe-etal-2020-unsupervised,samih-darwish-2021-topical} have also been proposed for classifying individuals' stances without the need for labelled data. These methods utilise linguistic features extracted from post content, such as lexicons, grammatical dependencies, and the latent meanings of words in context. Alternatively, they may leverage users' profile features to infer their stance. However, methods 
relying on linguistic features may encounter challenges in multilingual domains due to their dependency on language-specific characteristics~\cite{alturayeif2023systematic}. Meanwhile, those utilising profile features may face difficulties in generalising across diverse social media platforms, given the variability in profile features offered by each platform. 

Our method employs a lightweight unsupervised approach, that is language- and profile-agnostic, setting it apart from these previous studies. Moreover, it does not rely on user interactions, which may not always be available on certain social media platforms (e.g., users sharing opinions in a Telegram group). To our knowledge, Darwish et al.~\cite{10.1145/3110025.3110112} and Coletto et al.~\cite{10.1145/2911451.2914716}
are the only previous work addressing a classification task similar to our work. Darwish et al.~\cite{10.1145/3110025.3110112} propose an approach that relies on predefined sets of users, each representing a particular stance. The stance of each user is then determined based on their similarity to these predefined sets.
Although this method relies on keywords (hashtags) shared by users to detect stances, similar to ours, it differs in that it requires a set of users for each stance as labelled data, whereas our method relies on only one keyword for each stance. Coletto et al.~\cite{10.1145/2911451.2914716} propose an approach that classifies keywords and users iteratively. Similar to our approach, this method relies on the keywords shared by users and requires only one keyword for each stance as prior knowledge. However, this method requires pre-defined thresholds for classification, and determining the optimal thresholds for various domains is not trivial and, in some cases, not feasible. Additionally, our experimental evaluation demonstrates that our lightweight method outperforms both of these methods.  

The main contribution of this paper is the lightweight classification method, which utilises a tailored random walk approach to effectively classify users and hashtags with minimal user data and prior knowledge. We apply our method, called the Lightweight Random Walk Method~(LRM), to analyse conversations about the UK General Election 2019 and Brexit on Twitter (now X). Our approach shows superior performance compared to baseline methods in accurately classifying users into different stance classes and identifying the most prominent hashtags within each class. Our method also offers competitive running time compared to baseline methods. Additionally, the proposed model is utilised to track the evolution of stance classes over time in both the general election and Brexit datasets. Results obtained by tracking the evolution in the Brexit data are consistent with previous studies applying content and text analysis~\cite{grvcar2017stance,khatua2016leave}. 
Although there is a lack of previous studies on analysing stance in the 2019 general election, our work corroborates with reports~\cite{fletcher2019did} showing quantitative analysis about this election in social media. 


\section{Lightweight Random Walk Method~(LRM)}

Let $U= \{u_1, u_2, \cdots, u_n\}$ be a set of users sharing a set of hashtags $H=\{h_1,h_2, \cdots, h_m\}$, where $n$ and $m$ indicate the number of users and hashtags, respectively. Let $R_{n\times m}$ be the sharing records with $R_{ki}$ denoting the number of times $u_k$ shares $h_i$; $R_{ki}=0$ if $u_k$ does not share $h_i$. The notation $\mathcal{R}_i$ indicates the total number of times that hashtag $h_i$ is shared, calculated as the sum of sharing instances across all users, i.e., $\mathcal{R}_i = \sum_{k=1}^{n} R_{ki}$, and notation $\mathscr{R}_k$ represents the total number of times $u_k$ has shared hashtags, calculated as the sum of sharing instances of $u_k$ across all hashtags, i.e., $\mathscr{R}_k = \sum_{i=1}^{n} R_{ki}$.

Suppose there are $t$ stances (represented as $t$ classes) towards a controversial topic, with each stance $c$ (where $c \in \{1, \cdots, t\}$) associated with a given hashtag $s_c$, referred to as the seed hashtag or simply the seed. The objective is to classify hashtags and users into these $t$ classes using only one hashtag (seed) associated with each stance. 

The classification in the proposed method is carried out as follows: (1) \textit{Hashtag classification:} for each hashtag, we calculate the similarity between the hashtag and each seed. Then, the hashtag is assigned to class $c$ where $s_c$ has the highest similarity to the hashtag. The set of hashtags assigned to class $c$ is denoted by $\mathcal{C}_c$. \textit{User classification:} For each user, based on the class of the hashtags shared by the user, we determine the user's inclination towards each class. The user is assigned to class $c$ when they exhibit the highest inclination towards the class. The method comprises four steps described in the following sub-sections.



\subsection{Generation of a hashtag-sharing graph}

To construct the hashtag-sharing graph, sharing records are projected onto a graph where nodes represent hashtags and edges denote relationships between pairs of hashtags. This graph is defined using a matrix $A_{m\times m}$ where $A_{ij}$ denotes the strength of the relationship between hashtags $h_i$ and $h_j$. The value of $A_{ij}$ is computed using Eq.~(\ref{Eq_simple_graph_weight}).
 
\begin{equation}
    A_{ij}=\frac{\sum_{u_k\in U} {\min(R_{ki},R_{kj})}}{\min(\mathcal{R}_i,\mathcal{R}_j)}
    \label{Eq_simple_graph_weight}
\end{equation}

The value $A_{ij}$ falls within the range of $[0,1]$, with a larger value indicating a stronger relationship.


\subsection{Determination of similarity between hashtags and seeds}

We apply a modified local random walk algorithm to determine the similarity between each hashtag $h_i\in H$ and each seed $s_c$ (where $c\in\{1,\cdots,t\}$). In a local random walk algorithm, given a transition matrix $A$ (where $A_{ij}$ is regarded as the probability that a random walker will go to $h_j$ from $h_i$), a walker starts walking from one node and continues traversing the graph until a specified stopping criterion is met. The general equation of the random walk is as Eq.(\ref{Eq_RandomWalk1}):

\begin{equation}
    \pi_{c}(z)=A^T\cdot \pi_{c}(z-1)
    \label{Eq_RandomWalk1}
\end{equation}
where $\pi_c(z)$ is an $m\times 1$ vector in which the $i$-th element, denoted by $\pi_{ci}(z)$, represents the probability of visiting node $h_i$ in the $z$-th step of the walk. The notation $A^T$ indicates the transpose of the transition matrix, and $\pi_c(0)$ is a vector with all zero values except the $c$-th element, which is 1.

In our modified random walk algorithm, a walker starts from a seed hashtag $s_c$ and traverses the graph to calculate the similarity between each hashtag and the seed. However, to avoid assigning undue similarity between the given hashtag $s_c$ and other hashtags strongly related to other seed hashtags, the walker should be prevented from visiting hashtags with strong relationships to other seeds. This visiting may occur in two cases if a walker starting from $s_c$: (i) visits a different seed $s_j$ and then proceeds to hashtags strongly related to $s_j$, or (ii) visits a general hashtag, which is used by individuals from different stances (e.g., the hashtag {\it \#ge2019} in the context of the 2019 UK general election), and then proceeds to hashtags strongly related to other seeds. To address case (i), we set all the values in the corresponding row and column of $j\in \{1,\cdots,t\}-\{c\}$ in the transition matrix to zero to prevent the walker from visiting hashtags through seed $s_j$, ensuring that $s_c$ is the only seed involved in the walk. To address (ii), we utilise the notation of entropy to adjust the transition probabilities through every hashtag $h_i$, considering the generality of the hashtag across various stances. For this purpose, we first compute an entropy $E_i$, as shown in Eq.~(\ref{Eq_RelationtionEntropy}). The value of $E_i$ falls within the range of $[0,1]$, with a larger value indicating that the hashtag $h_i$ is more neutral. 
\begin{equation}
    E_{i}=\frac{-\sum_{c=1}^{t} \frac{A_{ic}}{\mathscr{A}_i}\cdot log(\frac{A_{ic}}{\mathscr{A}_i})}{log(t)}
    \label{Eq_RelationtionEntropy}
\end{equation}
In Eq.~(\ref{Eq_RelationtionEntropy}), $\mathscr{A}_i$ indicates the sum of the relationships between $h_i$ and all seed hashtags. Then, the value $A_{ic}$ is decreased as $A_{ic}=(1-E_i)/d(i)\cdot A_{ic}$, where $d_i$ is the degree of the node $h_i$ in the hashtag-sharing graph (the number of non-zero values in the $i$-th row of the matrix $A_{m\times m}$).

The random walk approach is iterated for each seed hashtag $s_c$, to calculate the similarity between hashtags and seeds. In each iteration, the matrix $A$ is row-normalised to ensure that the sum of values in each row is 1. Then, a walker traverses the graph by starting from a seed $s_c$. In a random walk approach, the similarity of every hashtag $h_i \in H$ to $s_c$ can be calculated using Eq. (\ref{Eq_RandomWalk2}).
\begin{equation}
    S_{ci}=\sum_{z=1}^{\rho} \pi_{ci}(z)
    \label{Eq_RandomWalk2}
\end{equation}
In this equation, $\pi_{ci}(z)$ represents the $i$-th element of vector $\pi_{c}(z)$, which can be calculated using Eq. (\ref{Eq_RandomWalk1}), and $\rho$ denotes the number of steps in the random walk approach. To balance between running time and classification performance, we empirically set this value to 10 in our experiments.


\subsection{Classification of Hashtags}
After determining the similarity between hashtags and all seeds, each hashtag $h_i \in H$ is assigned to class $c$ where the hashtag has the largest similarity value to the seed $s_c$. We use the notation $\mathcal{C}_c$ to refer to the set of hashtags assigned to class $c$. Additionally, we determine the stance intensity of each hashtag $h_i$ as shown in Eq. (\ref{EQ_HashtagPlarity}). 
\begin{equation}
    \mathcal{I}(h_i)=\frac{-\sum_{c=1}^{t} \frac{s_{ci}}{\mathscr{s}_i}\cdot log(\frac{s_{ci}}{\mathscr{s}_i}) }{log(t)}
    \label{EQ_HashtagPlarity}
\end{equation}
where $\mathscr{s}$ denotes the sum of the similarity values between $h_i$ and all seeds. The stance intensity $\mathcal{I}(h_i)$ falls within the range of $[0,1]$, where a larger value indicates a greater stance intensity.

\subsection{Classification of Users}

In this step, the stance and intensity of the hashtags used by each user $u_k$ are utilised to determine the user's inclination towards each class. The user is assigned to class $c$ where they exhibit the greatest inclination towards the hashtags in $\mathcal{C}_c$. The inclination of $u_k$ towards class $c$ is calculated using Eq. (\ref{EQ_UserLeaning}).
\begin{equation}
    \mathcal{L}_{kc}=\sum_{h_i \in \mathcal{C}_c} \mathcal{I}(h_i) \cdot \frac {R_{ki}}{\mathscr{R}_k}
    \label{EQ_UserLeaning}
\end{equation}
In this equation, $\mathscr{R}_k$ represents the total number of times $u_k$ has shared hashtags, and $R_{ki}$ represents the number of times user $u_k$ has shared hashtag $h_i$. The value of $\mathcal{L}_{kc}$ ranges    
 form 0 to 1, with a larger value indicating a stronger inclination of $u_k$ towards class $c$. 


\section{Experimental Evaluation}

\subsection{Datasets}
To assess the effectiveness of the proposed method, we employ two Twitter (X) datasets collected via Twitter Streaming APIs. For each dataset, initially, we compile a list of the most frequently used hashtags in the relevant context and collect tweets containing at least one of these hashtags.
\begin{enumerate}[label=(\roman*)]
\item The UK General Election Dataset (\textit{Election} dataset) covers the UK general election held on 12 December 2019. This dataset was collected over four weeks, from 16 November 2019 to 12 December 2019, and includes 234,800 unique hashtags shared by 929,441 unique accounts, totalling 14,233,501 sharing records. In this dataset, the five major political parties are treated as distinct classes: Labour~(\textit{LAB}), Conservative~(\textit{CON}), Liberal Democrats~(\textit{LDM}), Scottish National~(\textit{SCN}), and Green~(\textit{GRE}). For each party, a hashtag that explicitly represents the party's stance is selected as a seed. The chosen seeds are \textit{\#votelabour}, \textit{\#voteconservative}, \textit{\#votelibdem}, \textit{\#votesnp}, and \textit{\#votegreen} for the \textit{LAB}, \textit{CON}, \textit{LDM}, \textit{SCN}, and \textit{GRE} classes, respectively. 
\item Brexit Referendum Dataset (\textit{Brexit} dataset) pertains to the Brexit referendum, held on 23 June 2016, where the UK voted to decide whether to leave or remain in the European Union. This dataset was collected over four weeks, from 27 May 2016 to 23 June 2016. It includes 158,673 unique hashtags shared by 1,242,530 unique accounts, totalling 15,446,338 sharing records.
In this dataset, two classes are considered: Remain~(\textit{RMN}) and Leave~(\textit{LVE}). The seed hashtags chosen for this dataset are {\it \#voteremain} and {\it \#voteleave} for the \textit{RMN} and \textit{LVE} classes, respectively.
\end{enumerate}

For each dataset, we establish golden sets of users and hashtags. The golden set of users comprises accounts of members of parliament, candidates, or government ministers known for supporting each stance present in the collected dataset, acting as representatives of their respective classes. 
A hashtag qualifies as a golden hashtag for a class if it is shared at least 30 times by more than one account in that class and at least five times more frequently than by accounts in any other class. These criteria ensure the establishment of a reasonable golden hashtag set with a practical size. Note that the golden sets of users and hashtags are used exclusively for evaluation purposes. Table~\ref{Tbl_GodelnSets} summarises the number of golden users and hashtags identified for evaluation in each stance (class) and dataset. 
During the experiments on both datasets, hashtags with low engagement are filtered out by excluding those with sharing frequencies lower than the average frequency of all hashtags. Similarly, users with sharing records fewer than the average sharing records of all users are also removed.


\subsection{Setting}
We compare our proposed LRM method with the following five methods: (i) The Iterative Classification Method~(ICM)~\cite{10.1145/2911451.2914716} which uses iterative classification to determine the stance of hashtags and users. (ii) The Hashtag Similarity Method~(HSM)~\cite{10.1145/3110025.3110112}, originally designed for user classification using seed users, has been adapted in this paper due to the lack of seed users. It now classifies hashtags based on similarity to seed hashtags. (iii) Random Method~(RDM) that randomly assigns hashtags and users to classes. (iv) The Label Propagation Method~(LPM) which constructs a graph of hashtag co-occurrences and uses label propagation to assess the similarity between each hashtag and seeds. (v) Simple Random Walk~(SRM) that generates a graph of hashtag co-occurrences and applies a random walk to measure the similarity to the seeds. ICM uses iterative classification for both hashtags and users, while RDM assigns them randomly. HSM, LPM, and SRM classify hashtags based on similarity to seeds. For user classification, these methods assign users to the class with the most shared hashtags.

\begin{table}[h]
\centering
\caption{The golden sets (users and hashtags)}
\scriptsize{
\begin{tabular}{|c||c|c|c|c|c||c|c|}
\hline
Dataset & \multicolumn{5}{c||}{Election dataset} & \multicolumn{2}{c|}{Brexit dataset} \\
\hline
 Class & \textit{LAB} & \textit{CON} & \textit{LDM} & \textit{SCN} & \textit{GRE} & \textit{RMN} & \textit{LVE} \\
\hline
 User & 349 & 446 & 602 & 104 & 367 & 445 & 145 \\
\hline
 Hashtag & 80 & 109 & 53 & 29 & 32 & 36 & 28 \\
\hline
\end{tabular}
}
\label{Tbl_GodelnSets}
\end{table}

Three experiments were conducted for evaluation. The first assesses performance in classifying golden users and hashtags using $F_1$ scores for each class and $F_{Macro}$ score as an indicator of overall performance. The second evaluates method efficiency based on average running time. Lastly, the third uses the proposed method to track the evolution in the weeks leading up to the events (election/referendum).


\subsection{Results}

In the first experiment, each method is applied to classify the users and hashtags using the seed hashtags for both datasets; and evaluated using the golden users and hashtags sets. 
Table~\ref{Tbl_Generalelection_Hashtags_Users} shows the results in terms of $F_1$ and $F_{Macro}$ for the Election dataset, whilst Table~\ref{Tbl_Brexit_Hashtags_Users} summarises the results for the Brexit dataset.
The best performance in each class ($F_1$) and across all classes ($F_{Macro}$) is shown in \textbf{bold face} in these tables.


\begin{table}[h!]
\centering
\caption{Classification performance in the Election dataset}
\scriptsize{
\begin{tabular}{|c||c|c|c|c|c|c||c|c|c|c|c|c|}
\hline
&  \multicolumn{6}{c||}{Hashtag Classification}& \multicolumn{6}{c|}{User Classification} \\
 \cline{2-13}
Score & ICM & HSM & RDM & LPM & SRM & LRM & ICM & HSM & RDM & LPM & SRM & LRM \\
\hline
$F_1$ (LAB)	&	\textbf{0.919}	&	0.710	&	0.225	&	0.609	&	0.741	&	0.909	&	0.944	&	0.659	&	0.248	&	0.569	&	0.827	&	\textbf{0.975}	\\ \hline
$F_1$ (CON)	&	\textbf{0.913}	&	0.870	&	0.249	&	0.688	&	0.850	&	\textbf{0.913}	&	0.911	&	0.891	&	0.237	&	0.532	&	0.861	&	\textbf{0.969}	\\ \hline
$F_1$ (LDM)	&	0.901	&	0.516	&	0.191	&	0.351	&	0.739	&	\textbf{0.913}	&	0.831	&	0.367	&	0.198	&	0.140	&	0.811	&	\textbf{0.938}	\\ \hline
$F_1$ (SCN)	&	0.897	&	\textbf{0.981}	&	0.134	&	0.963	&	0.732	&	0.912	&	0.852	&	\textbf{0.944}	&	0.069	&	0.727	&	0.459	&	0.897	\\ \hline
$F_1$ (GRE)	&	0.867	&	0.595	&	0.138	&	0.514	&	0.639	&	\textbf{1.000}	&	0.826	&	0.022	&	0.173	&	0.011	&	0.765	&	\textbf{0.956}	\\ \hline
$F_{Macro}$ 	&	0.899	&	0.734	&	0.188	&	0.625	&	0.740	&	\textbf{0.929}	&	0.873	&	0.577	&	0.185	&	0.396	&	0.745	&	\textbf{0.947}	\\ \hline

\end{tabular}
}
\label{Tbl_Generalelection_Hashtags_Users}
\end{table}

\begin{table}[h!]
\centering
\caption{Classification performance in the Brexit dataset}
\scriptsize{
\begin{tabular}{|c||c|c|c|c|c|c||c|c|c|c|c|c|}
\hline
&  \multicolumn{6}{c||}{Hashtag Classification}& \multicolumn{6}{c|}{User Classification} \\
 \cline{2-13}
Score & ICM & HSM & RDM & LPM & SRM & LRM & ICM & HSM & RDM & LPM & SRM & LRM \\
\hline
$F_1$ (RMN)	&	0.909	&	0.750	&	0.541	&	0.293	&	0.933	&	\textbf{0.941}	&	0.907	&	0.580	&	0.617	&	0.071	&	0.905	&	\textbf{1.000}	\\ \hline
$F_1$ (LVE)	&	0.880	&	0.767	&	0.432	&	0.613	&	0.878	&	\textbf{0.917}	&	\textbf{1.000}	&	0.446	&	0.280	&	0.337	&	0.258	&	\textbf{1.000}	\\ \hline
$F_{Macro}$ 	&	0.895	&	0.758	&	0.487	&	0.453	&	0.906	&	\textbf{0.929}	&	0.954	&	0.513	&	0.448	&	0.204	&	0.582	&	\textbf{1.000}	\\ \hline

\end{tabular}
}
\label{Tbl_Brexit_Hashtags_Users}
\end{table}


\begin{table}[ht!]
\centering
\caption{Running time of the different methods for both datasets (in minutes)}
\scriptsize{
\begin{tabular}{|c|c|c|c|c|c|c|}
\hline
Dataset	&	ICM 	&	HSM 	&	RDM 	&	LPM 	&	SRM 	&	LRM 	\\ \hline
Election dataset	&	 144.03 	&	 20.18 	&	 02.20 	&	 57.23 	&	 38.51 	&	 26.02 	\\ \hline
Brexit dataset	&	60.08	&	10.51	&	01.15	&	44.15	&	33.01	&	12.10	\\ \hline

\end{tabular}
}
\label{Tbl_RunningTime}
\end{table}

\begin{figure}[ht!]
 \center{
 
    \subfloat[The Election dataset]
    {
       \centering
       \includegraphics[width=0.46\textwidth,keepaspectratio]{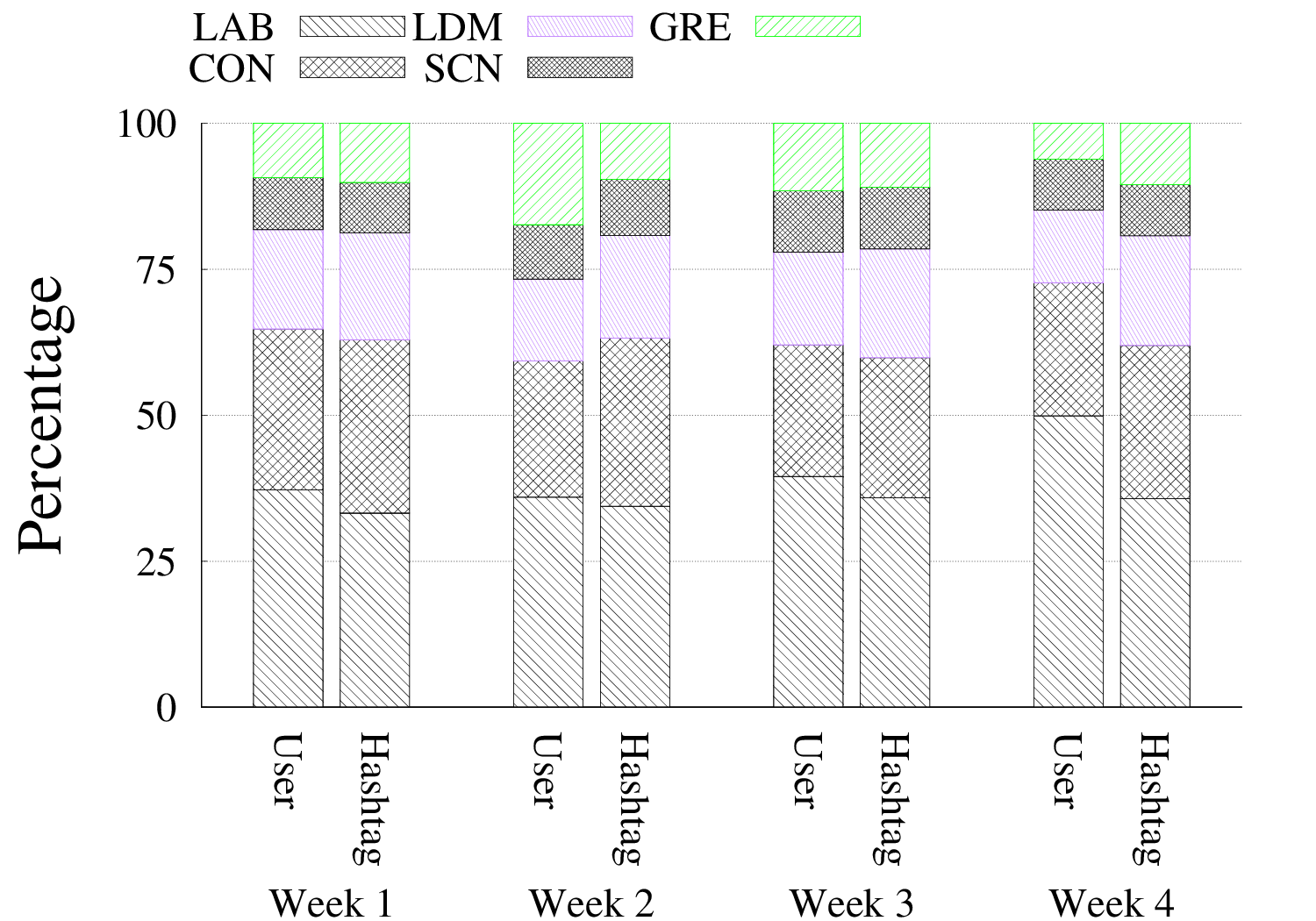}\label{InfluenceRelation}
    }
    \subfloat[The Brexit dataset]
    {
       \centering
       \includegraphics[width=0.46\textwidth]{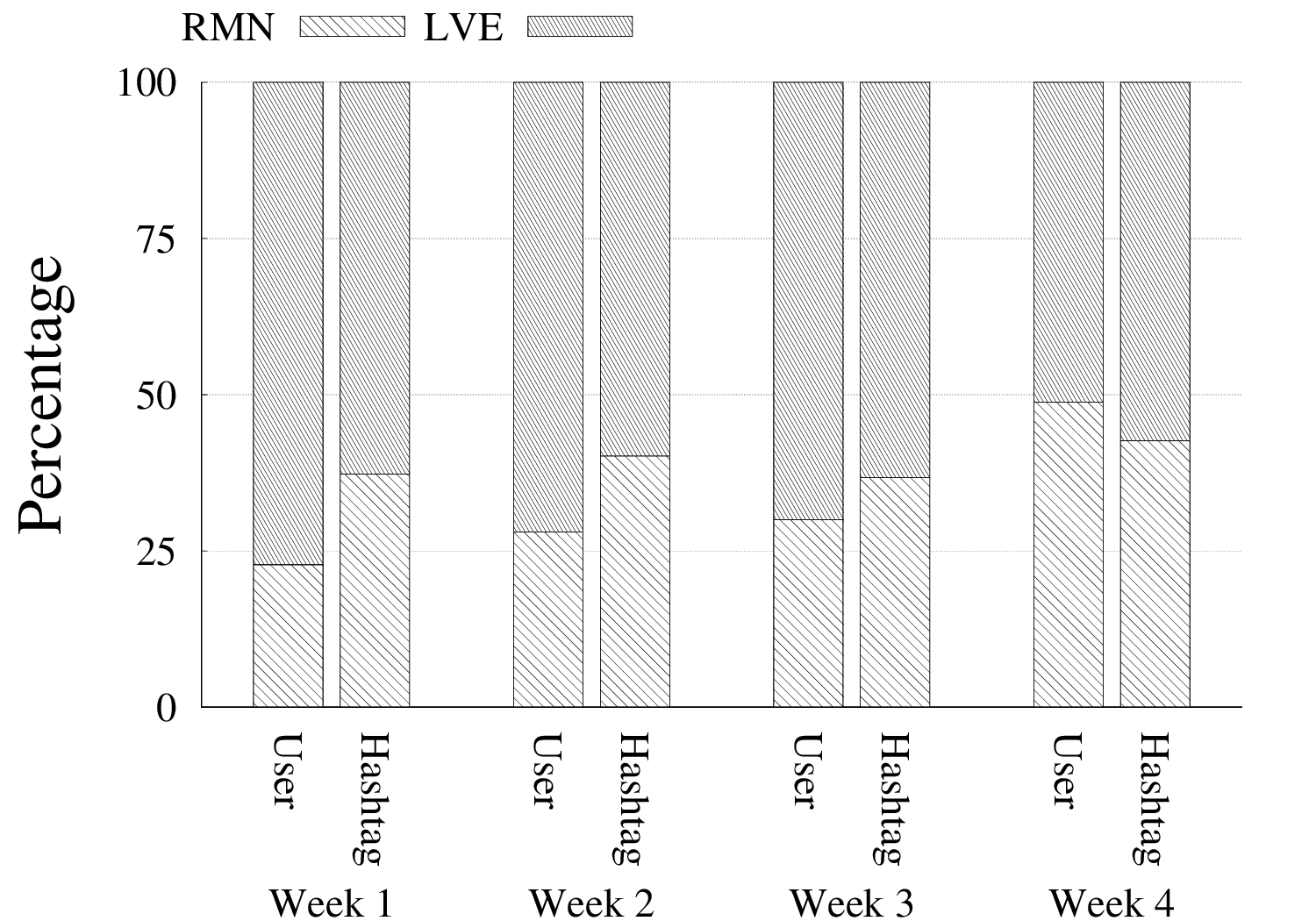}\label{ActivationRelation}
    }
   \caption{Percentage of users and hashtags in each class}
    \label{Fig_evolution}
    }
 \end{figure}

Regarding $F_{Macro}$ scores, 
the proposed method~(LRM) outperforms the other methods, with ICM providing the second-best results. Notably, SRM using a simple random walk also exhibits competitive performance, suggesting that employing the random walk approach is effective for calculating hashtag similarities for the task outlined in this paper. However, the modifications implemented in the random walk approach in LRM further enhance classification performance. The difference between the $F_{Macro}$ scores obtained by LRM and those of ICM (the second-best method) indicates that LRM exhibits greater superiority in user classification compared to hashtag classification. This observation underscores the effectiveness of our heuristic, which leverages hashtag stance intensity for user classification, enhancing performance. Regarding per-class $F_1$ scores, the LRM method consistently outperforms other methods, with few exceptions, where it shows comparable performance. The worst performance of LRM is in the Election dataset for users and hashtags in the \textit{SCN} class. This may be because \textit{SCN} is a relatively small class when compared to the other classes (see Table~\ref{Tbl_GodelnSets}), and the proposed method may not properly determine the similarity between the hashtags in this class and the associated seed hashtag.

The second experiment evaluates the efficiency of methods by measuring their running time (Table~\ref{Tbl_RunningTime}). LRM demonstrates superior efficiency compared to LPM, SRM, and ICM. This efficiency stems from LRM's approach of constructing the hashtag-sharing graph based solely on user hashtag-sharing frequencies, unlike LPM and SRM, which process all tweets to find hashtag co-occurrences for graph generation. The iterative processing of posts by ICM renders it the slowest method. While RDM and HSM are faster than LRM, they exhibit significantly lower classification performance, as observed in the first experiment.

In the third experiment, we employ LRM to analyse the evolution of the number of users and hashtags in each class over the four weeks leading up to the event (election/referendum). To conduct this analysis, we extract data collected in each week and utilise LRM to classify users and hashtags. Figure~\ref{Fig_evolution} illustrates the percentage of all users and hashtags in each class over the four weeks. In the Election dataset (Figure~\ref{InfluenceRelation}) it is observed that \textit{LAB} consistently maintains a higher percentage of users and hashtags compared to other classes, with \textit{CON} following closely. Interestingly, even in the week preceding the election (week 4), the percentage of \textit{LAB} users and hashtags remains higher than that of \textit{CON}. This observation aligns with reports suggesting greater engagement with Labour accounts on social media platforms~\cite{fletcher2019did}. In the Brexit dataset (Figure~\ref{ActivationRelation}), the activity of users supporting \textit{LVE} is higher than that of users advocating for \textit{RMN}. However, as the referendum date approaches, the percentage of \textit{RMN} users and hashtags increases. By week 4, the percentage of users advocating \textit{RMN} and \textit{LVE} is 48.80\% and 51.20\%, respectively, closely reflecting the outcome of the referendum and findings from research predicting the outcome based on Twitter post analysis~\cite{grvcar2017stance,khatua2016leave}. We further analyse the data specifically on the day of the referendum, 23 June 2019, to ascertain the percentage of users advocating each stance. The results indicate a higher percentage of users advocating for \textit{RMN} (59.59\%). 
This finding aligns with previous research, that shows a similar trend: a large engagement of the \textit{LVE} campaign since the early stages of the referendum process, with \textit{RMN} catching up at a later stage~\cite{grvcar2017stance}.


\section{Conclusion}
This paper introduced a novel lightweight method for understanding public opinions on controversial topics, discovering the hashtags used to advocate each stance, and tracking the evolution over time. The method requires only one hashtag representing each stance and relies solely on the hashtags shared by the users. Unlike methods that rely on text analysis techniques or user interactions, our approach is language-agnostic and can be applied across various social networks without requiring user interactions. It also aligns with a real-world scenario, where a social scientist or other stakeholders would be interested in monitoring a given event, while they can easily identify seed hashtags. Experimental results demonstrate that our method outperforms similar methods and baselines.

\section*{Acknowledgment}
This work is supported by the UK’s innovation agency (InnovateUK) grant number 10039039 (approved under the Horizon Europe Programme as VIGILANT, EU grant agreement number 101073921)\footnote{\url{https://www.vigilantproject.eu}}.

%
%
%
\bibliographystyle{splncs04}
\bibliography{Ref}

\begin{thebibliography}{10}
\providecommand{\url}[1]{\texttt{#1}}
\providecommand{\urlprefix}{URL }
\providecommand{\doi}[1]{https://doi.org/#1}

\bibitem{10.1145/3359307}
Aldayel, A., Magdy, W.: Your stance is exposed! analysing possible factors for stance detection on social media. Proceedings of the ACM on Human-Computer Interaction  \textbf{3} (2019)

\bibitem{ALDAYEL2021102597}
ALDayel, A., Magdy, W.: Stance detection on social media: State of the art and trends. Information Processing \& Management  \textbf{58}(4),  102597 (2021)

\bibitem{alturayeif2023systematic}
Alturayeif, N., Luqman, H., Ahmed, M.: A systematic review of machine learning techniques for stance detection and its applications. Neural Computing and Applications  \textbf{35}(7),  5113--5144 (2023)

\bibitem{10.1145/2911451.2914716}
Coletto, M., Lucchese, C., Orlando, S., Perego, R.: Polarized user and topic tracking in twitter. In: Proceedings of the 39th International Conference on Research and Development in Information Retrieval. p. 945–948 (2016)

\bibitem{10.1145/3110025.3110112}
Darwish, K., Magdy, W., Zanouda, T.: Improved stance prediction in a user similarity feature space. In: Proceedings of the 2017 IEEE/ACM International Conference on Advances in Social Networks Analysis and Mining. p. 145–148 (2017)

\bibitem{Darwish_Stefanov_Aupetit_Nakov_2020}
Darwish, K., Stefanov, P., Aupetit, M., Nakov, P.: Unsupervised user stance detection on twitter. Proceedings of the International AAAI Conference on Web and Social Media  \textbf{14}(1),  141--152 (2020)

\bibitem{fletcher2019did}
Fletcher, R.: Did the conservatives embrace social media in 2019. UK election analysis pp. 1--123 (2019)

\bibitem{grvcar2017stance}
Gr{\v{c}}ar, M., Cherepnalkoski, D., Mozeti{\v{c}}, I., Kralj~Novak, P.: Stance and influence of twitter users regarding the brexit referendum. Computational social networks  \textbf{4},  1--25 (2017)

\bibitem{khatua2016leave}
Khatua, A., Khatua, A.: Leave or remain? deciphering brexit deliberations on twitter. In: 2016 IEEE 16th international conference on data mining workshops. pp. 428--433 (2016)

\bibitem{kobbe-etal-2020-unsupervised}
Kobbe, J., Hulpu{\textcommabelow{s}}, I., Stuckenschmidt, H.: Unsupervised stance detection for arguments from consequences. In: Proceedings of the 2020 Conference on Empirical Methods in Natural Language Processing. pp. 50--60 (2020)

\bibitem{10.1145/3003433}
Mohammad, S.M., Sobhani, P., Kiritchenko, S.: Stance and sentiment in tweets. ACM Transactions on Internet Technology  \textbf{17}(3) (2017)

\bibitem{samih-darwish-2021-topical}
Samih, Y., Darwish, K.: A few topical tweets are enough for effective user stance detection. In: Proceedings of the 16th Conference of the European Chapter of the Association for Computational Linguistics: Main Volume. pp. 2637--2646 (2021)

\bibitem{10068608}
Williams, E.M., Carley, K.M.: Tspa: Efficient target-stance detection on twitter. In: 2022 IEEE/ACM International Conference on Advances in Social Networks Analysis and Mining. pp. 242--246 (2022)

\bibitem{zhou2023stance}
Zhou, L., Zhou, K., Liu, C.: Stance detection of user reviews on social network with integrated structural information. Journal of Intelligent \& Fuzzy Systems  \textbf{44}(2),  1703--1714 (2023)

\end{thebibliography}

\end{document}